\documentclass[a4paper, aps, amssymb, amsmath, reprint, showkeys, nofootinbib, twoside, superscriptaddress,longbibliography]{revtex4-2}
\usepackage[version=3]{mhchem} 
\usepackage{placeins}
\usepackage{tikz}
\usetikzlibrary{quantikz}
\usepackage{braket}
\usepackage{subcaption}
\usepackage{caption}
\usepackage{todonotes}
\newcommand{\sr}[1]{_{\mathrm{#1}}}
\usepackage{comment}
\usepackage{multirow}
\usepackage[unicode=true, colorlinks=true]{hyperref}
\usepackage[capitalise]{cleveref}

\begin{document}
\title[]{{Reducing Unitary Coupled Cluster Circuit Depth by Classical Stochastic Amplitude Pre-Screening}}

\author{Maria-Andreea Filip}
    \email{maf63@cam.ac.uk}
    \affiliation{Yusuf Hamied Department of Chemistry\\Lensfield Road, CB2 1EW Cambridge\\ United Kingdom}
\author{Nathan Fitzpatrick}
    \email{nathan.fitzpatrick@cambridgequantum.com}
    \affiliation{Cambridge Quantum Computing Ltd.\\ 13-15 Hills Road, CB2 1NL Cambridge\\ United Kingdom}
\author{David Mu\~noz Ramo}
    \affiliation{Cambridge Quantum Computing Ltd.\\ 13-15 Hills Road, CB2 1NL Cambridge\\ United Kingdom}
\author{Alex J. W. Thom}
    \email{ajwt3@cam.ac.uk}
    \affiliation{Yusuf Hamied Department of Chemistry\\Lensfield Road, CB2 1EW Cambridge\\ United Kingdom}

\date{\today}
\begin{abstract}
Unitary Coupled Cluster (UCC) approaches are an appealing route to utilising quantum hardware to perform quantum chemistry calculations, as quantum computers can in principle perform UCC calculations in a polynomially scaling fashion, as compared to the exponential scaling required on classical computers.
Current noisy intermediate scale quantum (NISQ) computers are limited by both hardware capacity in number of logical qubits and the noise introduced by the deep circuits required for UCC calculations using the Variational Quantum Eigensolver (VQE) approach.
We present a combined classical--quantum approach where a stochastic classical UCC pre-processing step is used to determine the important excitations in the UCC ansatz.
The reduced number of selected excitations are then used in a UCC-based VQE calculation.
This approach gives a systematically improvable approximation, and we show that significant reductions in quantum resources can be achieved, {with simulations on the \ce{CH2}, \ce{N2} and \ce{N2H2} molecules giving sub-milliHartree errors.}
\end{abstract}
\maketitle

\section{Introduction}
Quantum chemistry, which often concerns itself with solutions to problems with exponentially scaling Hilbert spaces, has long been identified as a good target for quantum computation, which would allow the encoding of such problems in a linear number of qubits. For example, algorithms such as Quantum Phase Estimation\cite{Abrams1997, Abrams1999} (QPE) have been suggested as means of efficiently computing the ground state energies of molecular systems.\cite{Aspuru-Guzik2005} However, this approach would require fault-tolerant quantum computers. In the current noisy intermediate-scale quantum (NISQ) regime, the field of quantum chemistry calculations on quantum computers is dominated by hybrid quantum-classical approaches and in particular the Variational Quantum Eigensolver\cite{Peruzzo2014} (VQE), which requires a parameterised wavefunction. Various parameterised ansatz schemes have been proposed, such as the hardware efficient ansatz \cite{Kandala2017}, the hardware variational ansatz\cite{PRXQuantum.1.020319}, the symmetry preserving ansatz \cite{Gard2020} and the unitary coupled cluster (UCC) ansatz \cite{Kutzelnigg1982, Kutzelnigg1983,Kutzelnigg1984,Bartlett1989}. UCC has seen a resurgence as a convenient parameterisation for VQE,\cite{Peruzzo2014, Barkoutsos2018} due to its ability to be easily encoded into a quantum circuit. While its physical origin leads to a series of appealing features, including a relatively well-behaved energy landscape, the qubit encoding of the fermionic operators in the UCC ansatz produces very deep circuits, which are challenging for currently available quantum architectures. Work has been done to decrease the depth of these circuits while maintaining the physicality of the ansatz, for example by using Moller-Plesset\cite{Moller1934} (MP2) theory results to screen amplitudes\cite{Romero2018, metcalf_ducc} or using adaptive ans\"atze such as ADAPT-VQE.\cite{Grimsley2019b} While the latter does generate shorter, highly accurate ans\"atze, it does so at the cost of an increased number of measurements relative to the standard UCC approach. {Other approaches which select contributing amplitudes based on their energy gradient have been developed, such as Qubit Coupled Cluster (QCC)\cite{Ryabinkin2018, Ryabinkin2020} or energy-sorted UCC,\cite{Fan2021} but like ADAPT-VQE they require additional quantum computation to obtain the screened set of operators.}

In this paper, we propose using the newly developed unitary coupled cluster Monte Carlo\cite{Filip2020} (UCCMC) approach to screen amplitudes for a UCC-based VQE calculation. Quantum Monte Carlo (QMC) methods\cite{Booth2009,Thom2010} take advantage of the sparsity of most chemical Hamiltonians to generate compact wavefunctions, thereby lowering memory requirements relative to the corresponding classical algorithms. Additionally, the Monte Carlo approach naturally samples ``important" contributions to the wavefunction --- determinants with large Hamiltonian coupling terms to the reference determinant --- first and therefore, while fully converging a QMC calculation may be time-consuming, short runs may be used to quickly identify the most important contributions to a given wavefunction.\cite{Deustua2017} We use this property of the UCCMC method and in particular its trotterized approximation to provide an initial set of amplitudes for a VQE calculation. We screen amplitudes based on these initial values and assess the effect of using the screened parameter sets on the accuracy of the obtained energy, comparing with the equivalent result from MP2 screening.

In Section 2, we review the underlying theory of both UCC-based VQE and the UCCMC method. Section 3 comprises results and discussion for a range of small molecules in a variety of scenarios and Section 4 presents our conclusions.

\section{Theory}
Coupled cluster theory \cite{Cizek1966, Cizek1969} has become well established as the ``gold-standard'' of \textit{ab initio} quantum chemistry methods. The exponential ansatz
\begin{equation}
    \ket{\Psi_\mathrm{CC}} = e^{\hat T}\ket{\Psi_0},
    \label{eq:wfn-cc}
\end{equation}
where $\hat T = \sum_{i} \hat T_i$ and $\hat T_i$ is composed of all valid excitation operators of order $i$, naturally maintains size-consistency when the operator $\hat T$ is truncated at some excitation level, with the most commonly employed truncations being coupled cluster singles and doubles (CCSD) and coupled cluster singles, doubles and triples (CCSDT). Such methods are polynomially scaling with system size and systematically improvable, both highly desirable properties. However, the exponential operator in this form is non-unitary and therefore it cannot be directly implemented on a quantum computer.

\subsection{The UCC ansatz}

We can construct the anti-Hermitian operator $\hat T - \hat T^\dagger$, the exponential of which gives a unitary operator. The UCC wavefunction is therefore

\begin{equation}
    |\Psi_\mathrm{UCC} \rangle = e^{\hat{T}-\hat{T}^\dagger}|\Psi_0\rangle
\end{equation}
and its energy can be found variationally as

\begin{equation}
E_0 = \min_\mathbf{t} \langle \Psi_0 | e^{-({\hat{T}-\hat{T}^\dagger})} \hat{H} e^{\hat{T}-\hat{T}^\dagger} |\Psi_0\rangle
\end{equation}

Although $e^{-({\hat{T}-\hat{T}^\dagger})} \hat{H} e^{\hat{T}-\hat{T}^\dagger}$ has a non terminating Baker--Campbell--Hausdorff (BCH) expansion, which makes the implementation of UCC costly on a classical computer, the unitary operator $e^{\hat{T}-\hat{T}^\dagger}$ can be decomposed as a series of universal quantum gates on a quantum computer. We start by writing the cluster operator in terms of individual excitations.

\begin{equation}
    U(\mathbf{t}) = e^{\sum_\mathbf{n} t_\mathbf{n} (\tau_\mathbf{n} - \tau_\mathbf{n}^\dagger)}
\end{equation}
where $\tau_\mathbf{n}$ represents a fermionic excitation operator and $t_\mathbf{n}$ its corresponding cluster amplitude. We can take a Trotter-Suzuki expansion\cite{Trotter1959, Suzuki1976} of $U(\mathbf{t})$ such that

\begin{equation}
|\Psi_\mathrm{UCC} \rangle \approx |\Psi_\mathrm{tUCC} \rangle = \Bigg ( \prod_\mathbf{n} e^{\frac{t_\mathbf{n}}{\rho}(\tau_\mathbf{n} - \tau_\mathbf{n}^\dagger)} \Bigg)^\rho |\Psi_0\rangle 
\end{equation}
with equality achieved as $\rho \rightarrow \infty$. We can take $\rho = 1$ which has been shown\cite{Barkoutsos2018} to be a good approximation, leading to negligible errors in the resulting energy, to obtain

\begin{equation}
\ket{\Psi_\mathrm{tUCC}} \approx U_{1}(\mathbf{t}) |\Psi_0\rangle = \prod_\mathbf{n} e^{t_\mathbf{n}(\tau_\mathbf{n} - \tau_\mathbf{n}^\dagger)} |\Psi_0\rangle,
\end{equation}
also known as the disentangled UCC ansatz. \cite{Evangelista2019}For the case of UCCSD the singles are represented by

\begin{equation}
\hat T_1 = \sum_{i,\alpha} t_{i}^{\alpha}(\hat a^\dagger_i \hat a_\alpha - \hat a^\dagger_\alpha \hat a_i),
\end{equation}
and doubles are represented by

\begin{equation}
\hat T_2 = \sum_{i>j,\alpha > \beta} t_{ij}^{\alpha\beta}(\hat a^\dagger_\alpha \hat a^\dagger_\beta \hat a_i \hat a_j - \hat a^\dagger_j \hat a^\dagger_i \hat a_\alpha \hat a_\beta),
\end{equation}
where the Latin indices represent occupied spin orbitals and the Greek indices represent virtual spin orbitals. This can be easily implemented on a quantum computer via an appropriate transformation of the fermionic operators to the unitary bosonic qubit operators using schemes like Jordan-Wigner\cite{Jordan1928} or Bravyi-Kitaev\cite{Bravyi2002}, where algebraic compilation strategies based on the ZX calculus have been proposed \cite{cowtan2020generic}.

\subsection{Quantum computing and the UCC ansatz}

NISQ hardware is characterized by small qubit counts and high noise levels, due mainly to short qubit coherence times, quantum gate errors, faulty readout operations and cross-talk between qubits during operation. These limitations prevent the use of error-correction protocols that would enable large quantum computations. Thus, it is advantageous to reduce the amount of quantum computation to a minimum. As a result, there is a strong research effort dedicated to the development of hybrid quantum--classical algorithms. One of the main algorithmic workhorses of this field is the Variational Quantum Eigensolver\cite{Peruzzo2014} , where a minimizer run on a classical machine optimizes a cost function evaluated by the quantum computer. In chemistry problems, this corresponds to finding the expected value of the energy with respect to a parametrized ansatz wavefunction. Alternative approaches have appeared in recent years; we highlight methods based on imaginary time propagation of a trial wavefunction, either via a variational principle\cite{McArdle2019} or approximating the nonunitary evolution with an appropriate quantum circuit \cite{Motta2019}, and techniques to variationally optimise the reduced density matrix of the system \cite{mazziotti2021rdm}.

\begin{figure*}[]
\centering
\begin{quantikz}
\lstick{$|i_3\rangle$} & \gate{R_x(\frac{3\pi}{2})} & \ctrl{1} & \qw & \qw &\qw & \qw & \qw & \ctrl{1} &\gate{R_x(\frac{\pi}{2})} & \qw \\
\lstick{$|i_2\rangle$} & \gate{H} &\targ{}  & \ctrl{1} & \qw & \qw & \qw & \ctrl{1}  & \targ{} & \gate{H} & \qw \\
\lstick{$|i_1\rangle$} & \gate{H} &\qw & \targ{}  & \ctrl{1} & \qw & \ctrl{1} & \targ{} & \qw & \gate{H} & \qw \\
\lstick{$|i_0\rangle$} & \qw &\qw & \qw & \targ{}  & \gate{R_z(\theta)} & \targ{}  & \qw & \qw & \qw & \qw \\
\end{quantikz}
\caption{\small Pauli gadget for $e^{-i\frac{\theta}{2}(Z_0 \otimes X_1 \otimes X_2 \otimes Y_3)}$}
\label{gadget2}
\end{figure*}
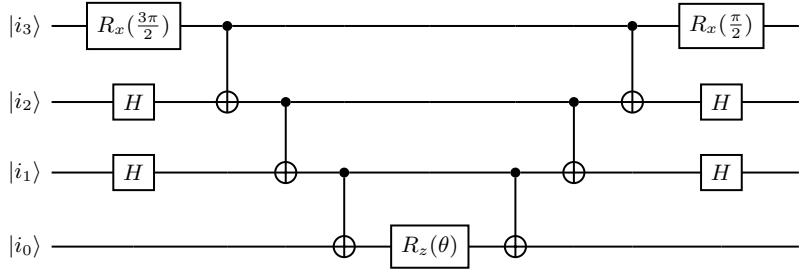

The UCC ansatz is attractive for this kind of quantum algorithm as it has a reduced number of parameters to optimize and a more favourable energy-search landscape compared with hardware-tailored ans\"{a}tze.\cite{Kandala2017} It also naturally conserves the number of electrons and the $M_z$ spin projection quantum number during the calculation, thus helping to prevent convergence to unwanted states, or avoiding the barren plateau problem observed for hardware-efficient ans\"{a}tze.\cite{Cerezo2021} However, these advantages are obtained at the cost of large circuit depths. For example, in the Jordan-Wigner mapping, fermionic creation and annihilation operators may be expressed as
\begin{equation}
a^\dagger_j = \bigotimes^{j-1}_i Z_i \bigotimes \frac{1}{2}(X_j - iY_j),
\end{equation}

\begin{equation}
a_j = \bigotimes^{j-1}_i Z_i \bigotimes \frac{1}{2}(X_j + iY_j).
\end{equation}
Therefore a single fermionic excitation operator becomes
\begin{equation}
t_{i}^{\alpha}(a^\dagger_i a_a - a^\dagger_a a_i) = \frac{it_{i}^{\alpha}}{2} \bigotimes_{k=i+1}^{\alpha-1} Z_k(Y_iX_\alpha - X_iY_\alpha),
\end{equation}
while a double excitation is given by 
\begin{widetext}
\begin{align}
\begin{split}
t_{ij}^{\alpha\beta}(a^\dagger_\alpha a^\dagger_\beta a_i a_j - a^\dagger_j a^\dagger_i a_\alpha a_\beta) = \frac{it_{ij}^{\alpha\beta}}{8}  \bigotimes_{k=i+1}^{j-1} Z_k  \bigotimes_{l=\alpha+1}^{\beta-1} Z_l (&X_i X_j Y_\alpha X_\beta + Y_i X_j Y_\alpha Y_\beta +X_i Y_j Y_\alpha Y_\beta + X_i X_j X_\alpha Y_\beta \\
& - Y_i X_j X_\alpha X_\beta - X_i Y_j X_\alpha X_\beta - Y_i Y_j Y_\alpha X_\beta - Y_i Y_j X_\alpha Y_\beta ).
\end{split}
\end{align}
\end{widetext}
Therefore, each excitation included in the ansatz contributes with a series of Pauli gadgets (see \cref{gadget2}), requiring a large number of two-qubit gates. These gates are responsible for most of the noise produced during a quantum computation on NISQ devices, so it is very important to reduce their number. Being able to identify which excitations have a negligible contribution to the total energy would enable their elimination in the ansatz and reduce the overall circuit depth of the state preparation step in variational quantum algorithms. Successful screening of excitations aims to bring the circuit execution time within the coherence time of near term hardware with maximal fidelity with the unscreened ansatz, incurring a minimum energy penalty.

\subsection{QMC algorithms}

Monte Carlo algorithms have become popular approaches to more effectively use classical computational resources for quantum chemistry. Real-space approaches such as Diffusion Monte Carlo (DMC)\cite{Anderson1975,Ceperley1986} suffer from the so-called ``sign problem", which causes them to naturally converge to bosonic solutions. This can be somewhat mitigated by using arbitrary nodal surfaces, however this introduces an uncontrolled approximation. Recently, Hilbert space quantum Monte Carlo methods have been developed that naturally avoid the DMC sign problem. Full Configuration Quantum Monte Carlo (FCIQMC)\cite{Booth2009} encodes a stochastic solution to the full configuration interaction (FCI) equation. The FCI wavefunction is expressed as a linear combination of a reference determinant (usually the Hartree--Fock (HF) wavefunction) and all possible excited determinants starting from it.

\begin{equation}
    \ket{\Psi_\mathrm{FCI}} = (1 + \hat C)\ket{D_0}
\end{equation}
where $\hat C = \sum_{i,\alpha} C_i^\alpha \hat a_i^\alpha + \frac{1}{4}\sum_{i,j,\alpha,\beta}C_{ij}^{\alpha\beta} \hat a_{ij}^{\alpha\beta} + ... $ and $\hat a_i^\alpha, \hat a_{ij}^{\alpha\beta}$ are excitation operators.
The coefficients $\mathbf{C}$ can be optimised by minimising the energy with respect to them. This gives the following set of equations:
\begin{equation}
    \braket{D_\mathbf{i}|\hat H - E|\Psi_\mathrm{FCI}}  = 0
    \label{eq:fci}
\end{equation}
where $\ket{D_\mathbf{i}}$ span the full Hilbert space of the system. Equivalently,
\begin{equation}
    \braket{D_\mathbf{i}|1-\delta\tau (\hat H - E)|\Psi_\mathrm{FCI}}  = \braket{D_\mathbf{i}|\Psi_\mathrm{FCI}}
\end{equation}
which can be written in an iterative form as
\begin{widetext}
\begin{equation}
    C_\mathbf{i}(\tau) - \delta\tau \braket{D_\mathbf{i}|\hat H -E|D_\mathbf{i}}C_\mathbf{i}(\tau) - \sum_\mathbf{j} \delta\tau\braket{D_\mathbf{i}|\hat H|D_\mathbf{j}}C_\mathbf{j}(\tau) = C_\mathbf{i}(\tau + \delta \tau)
\end{equation}
\end{widetext}
This equation can be solved stochastically by sampling the population dynamics of a set of walkers (`psips') in the Hilbert space of the system, which undergo the following processes:\cite{Booth2009}
\begin{itemize}
    \item spawning from $\ket{D_\mathbf{i}}$ to $\ket{D_\mathbf j}$ with probability 
\begin{equation}
p_\mathrm{spawn}(\mathbf{j}|\mathbf{i}) \propto \delta \tau|H_\mathbf{ij}|;
\end{equation}
\item death with probability 
\begin{equation}
p_\mathrm{death}(\mathbf{i}) \propto \delta \tau |H_
\mathbf{ii} - S|;
\end{equation}
    \item annihilation of particles of opposite sign on the same determinant.
\end{itemize}
In the death step, the shift $S$ replaces the unknown exact energy $E$. This acts as a population control parameter and, once the system has reached a steady state, converges to the true energy. Another estimator for $E$ is the projected energy 
\begin{equation}
    E_\mathrm{proj} = \frac{\braket{D_0|\hat H|\Psi}}{\braket{D_0|\Psi}} = \sum_{\mathbf{i} \neq 0} \frac{N_\mathbf{i}^\mathrm{CI} H_{\mathbf{i}0}}{N_0}
\end{equation}

A set of equations similar to Eq. \ref{eq:fci} holds for the CC wavefunction given in Eq. \ref{eq:wfn-cc}. Since $\braket{D_\mathbf{i}|\Psi_\mathrm{CC}} = t_\mathbf{i} + \mathcal{O}(t^2)$, we can write
\begin{equation}
    t_\mathbf{i}(\tau) - \delta\tau \braket{D_\mathbf{i}|\hat H -E|\Psi_\mathrm{CC}} \approx t_\mathbf{i}(\tau + \delta \tau).
    \label{eq:pop_dyn}
\end{equation}
This can be described stochastically by the same three processes considered for FCIQMC, leading to an algorithm know as Coupled Cluster Monte Carlo (CCMC).\cite{Thom2010} However, one must also take into consideration contributions from composite clusters.
For example,
\begin{equation}
    \braket{D_{ij}^{\alpha\beta}|\Psi_\mathrm{CC}} = t_{ij}^{\alpha\beta} + t_i^\alpha t_j^\beta - t_i^\beta t_j^\alpha,
\end{equation}
and therefore any of these three terms may contribute to death on $t_{ij}^{\alpha\beta}$ or to spawning onto some excitor coupled to it by the Hamiltonian. The selection process is therefore somewhat more complicated than for FCIQMC, originally consisting of the steps below:\cite{Thom2010}
\begin{enumerate}
\item a cluster size $s$ is selected with probability 
\begin{equation}
p(s) = \frac{1}{2^{s+1}}
\end{equation}

\item a particular cluster of $s$ distinct excitors is selected with probability 
\begin{equation}
p(e|s) = s! \prod_{i=1}^s \frac{|N_i|}{|N_\mathrm{ex}|}
\label{eq:psel}
\end{equation}
where $N_\mathrm{ex}$ is the total population on excitors. The total selection 
probability is therefore 
\begin{equation}
p_\mathrm{sel}(e) = p(e|s)p(s)
\end{equation}

\end{enumerate}
Improvements have since been made to this selection algorithm to better importance-sample the wavefunction.\cite{Scott2017}

Recently, some of us have implemented a stochastic version of unitary coupled cluster and its trotterized approximation,\cite{Filip2020} the details of which are expanded upon in the following section.

\subsubsection{(p)UCCMC}
While the UCC wavefunction is generally found by variationally optimising the parameters, it is also possible\cite{Pal1984, Evangelista2011} to solve a set of projected UCC (pUCC) equations, 
\begin{equation}
\braket{D_\mathbf{i}|\hat H - E|\Psi_\mathrm{UCC}} = 0.
\end{equation}
This projective method is naturally more amenable to the QMC algorithms described above, as it leads to similar population dynamics to those in Eq. \ref{eq:pop_dyn}. However,
the presence of de-excitation operators $\hat T^\dagger$ in the full UCC
ansatz substantially changes the structure of the allowed clusters, removing the constraint that cluster sizes must be at most equal to the maximum excitation level considered in the calculation. However, the expansion can be truncated
to a finite cluster size, which, if large enough,
does not significantly affect the accuracy of the obtained answer.
The selection scheme for UCCMC is then as follows:
\begin{enumerate}
\item select a cluster size $s$ with probability \mbox{$p(s) = \frac{1}{2^{s+1}}$}. 
\item for all but the first excitor in the cluster, decide with probability 
$\frac{1}{2}$ whether it will be an excitation or de-excitation operator
\item a particular cluster is selected as before, with probability given by 
\cref{eq:psel}
\end{enumerate}

Having selected the cluster, it undergoes stochastic spawning and death as before.  Overall this constitutes the projected Unitary Coupled Cluster Monte Carlo (pUCCMC) algorithm.
The final aspect one must be careful of is the projection of the wavefunction onto the HF reference, as this includes contributions beyond the reference population. This projection may be sampled stochastically during the course of the calculation, concurrently with the reference population and the $\frac{N_\mathbf{i}^\mathrm{CI} H_{\mathbf{i}0}}{N_0}$ terms.

\subsubsection{Trotterized pUCCMC}

Consider once again the trotterized UCC ansatz with $\rho = 1$. 
\begin{equation}
\ket{\Psi_\mathrm{tUCC}} = \prod_\mathbf{n} e^{t_\mathbf{n}(\tau_\mathbf{n} - \tau_\mathbf{n}^\dagger)} |\Psi_0\rangle
\end{equation}
This ansatz now depends on the order of excitors in the product, with different orders leading to different cluster amplitudes and potentially different energy values.\cite{Grimsley2019} Recent work by Evangelista \textit{et al.}\cite{Evangelista2019} has defined an optimal ordering that guarantees complete wavefunction expressibility in this framework.
Applying $e^{t_\mathbf{i}(\tau_\mathbf{i} - \tau_\mathbf{i}^\dagger)}$ to an arbitrary single determinant 
wavefunction $\ket{\Psi}$ leads to three possibilities:
\begin{enumerate}
    \item $\hat \tau_\mathbf{i}^\dagger\ket{\Psi} = 0$ and $\hat \tau_\mathbf{i}\ket{\Psi} \neq 0$ 
\begin{equation}
e^{t_\mathbf{i}(\tau_\mathbf{i} - \tau_\mathbf{i}^\dagger)}\ket{\Psi} = \cos(t_\mathbf i)\ket{\Psi} + \sin(t_\mathbf i) \ket{\Psi_\mathbf{i}} 
\end{equation}
where $\ket{\Psi_\mathbf{i}}$ is the result of applying the excitation to $\ket{\Psi}$.
\item $\hat \tau_\mathbf{i}^\dagger\ket{\Psi} \neq 0$ and $\hat \tau_\mathbf{i}\ket{\Psi} = 0$ 
\begin{equation}
e^{t_\mathbf{i}(\tau_\mathbf{i} - \tau_\mathbf{i}^\dagger)}\ket{\Psi} = \cos(t_\mathbf i)\ket{\Psi} - \sin(t_\mathbf i) \ket{\Psi^\mathbf{i}} 
\end{equation}
where $\ket{\Psi^\mathbf{i}}$ is the result of applying the deexcitation to $
\ket{\Psi}$.
\item  $\hat \tau_\mathbf{i}^\dagger\ket{\Psi} = 0$ and $\hat \tau_\mathbf{i}\ket{\Psi} = 0$
\begin{equation}
e^{t_\mathbf{i}(\tau_\mathbf{i} - \tau_\mathbf{i}^\dagger)}\ket{\Psi} = \ket{\Psi}  
\end{equation}

\end{enumerate}
To translate this into a stochastic algorithm, for each excitor present in the 
wavefunction, the algorithm assesses which of the cases listed above is 
appropriate. If the excitor cannot be applied, the next excitor is 
considered. If the excitor can be applied, this is done with probability
\begin{equation}
p_\mathrm{excit} = \frac{|\sin(t)|}{|\sin(t)| + |\cos(t)|} 
\end{equation}
and the cluster amplitude is multiplied by $\pm \sin(t)$. With probability 
$1-p_\mathrm{excit}$, the operator is not applied and the cluster amplitude
is multiplied by $\cos(t)$. The cluster then undergoes the same spawning,
death and annihilation steps as in traditional CCMC, leading to a trotterized pUCCMC approach (tpUCCMC). As in 
the case of full pUCCMC, this ansatz modifies the projection of the wavefunction
onto the reference, making it different from the reference population. Depending on the
ordering of the excitors, a closed form for this projection may be found, but in general
it can easily be sampled during the stochastic propagation.

\subsection{Technical details}

The VQE calculations have been performed using a minimal (STO-3G) basis set for all molecular species considered. Molecular integrals and molecular orbital coefficients have been obtained using the PySCF package.\cite{pyscf} The wavefunction anstaz has been encoded into quantum circuits on a qubit register using the Jordan-Wigner scheme in the EUMEN program. For the optimization loop in the VQE procedure, the L-BFGS method has been applied. The state vector simulator Qulacs\cite{suzuki2020qulacs} has been used to evaluate the quantum circuits.

All QMC calculations have been carried out in a development version of HANDE-QMC.\cite{HANDE2018} Where directly compared, VQE and tpUCCMC calculations use the same ordering of excitors, applying all single (de)excitation operators ahead of the doubles. {For triplet states, restricted open-shell HF (ROHF) reference states were used for all Monte Carlo, MP2 and VQE calculations.} All screened VQE UCCSD calculations used the relevant {tpUCCMCSD/MP2} amplitudes as starting values for the parameters. 

\section{Results}

\begin{figure}[h]
\begin{subfigure}{0.5\textwidth}
    \includegraphics[width=\textwidth]{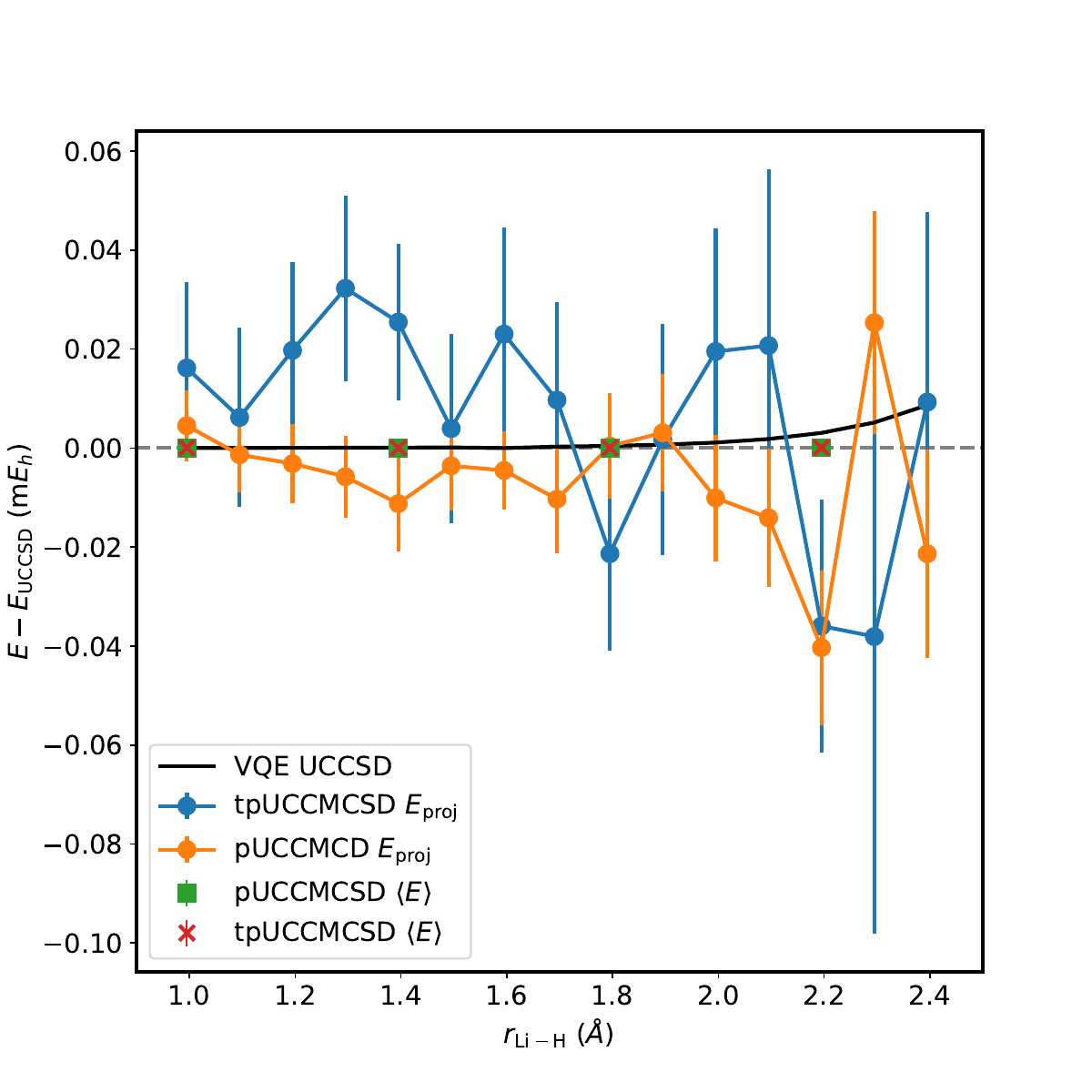} 
\end{subfigure}%
\begin{subfigure}{0.5\textwidth}
    \includegraphics[width=\textwidth]{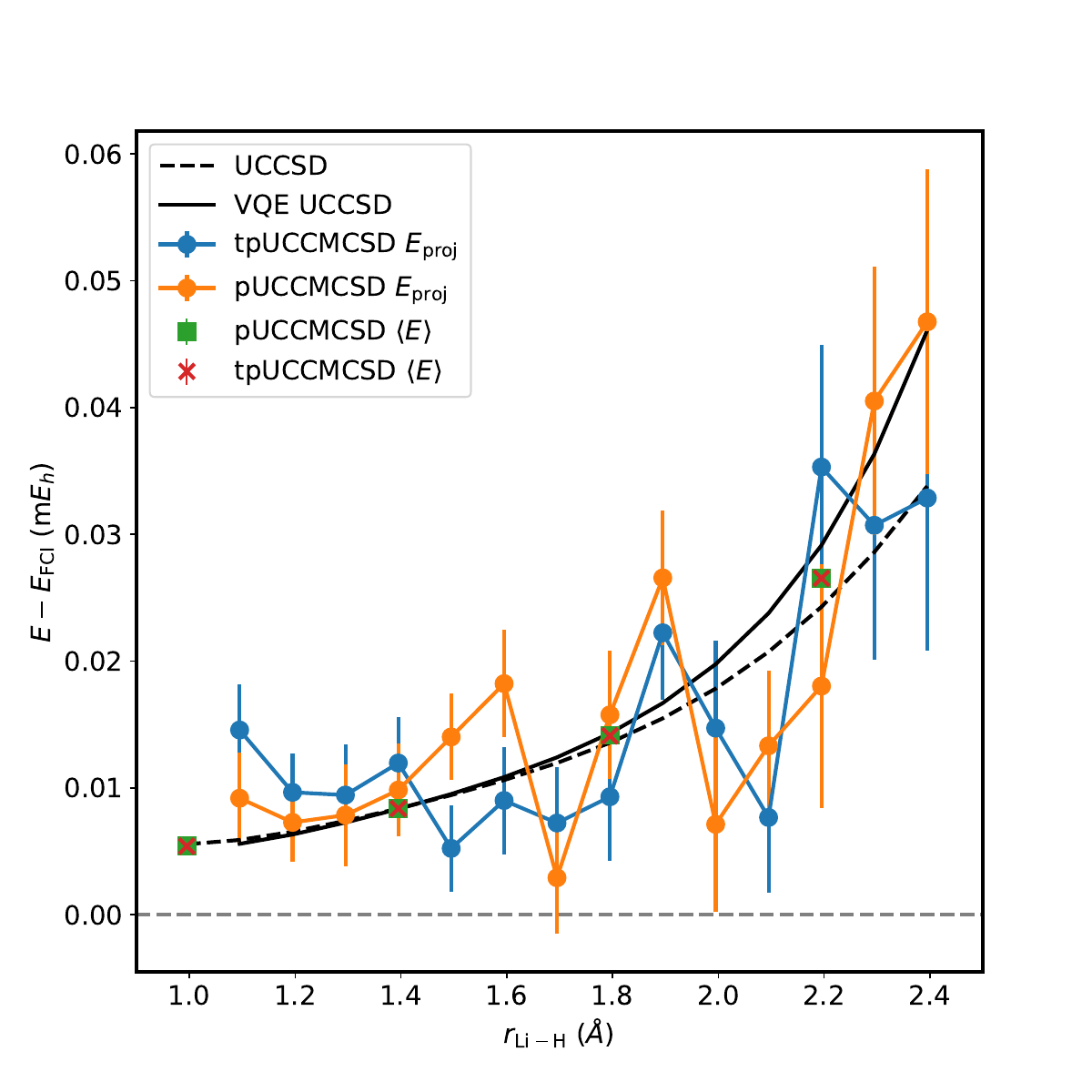} 
\end{subfigure}
    \caption{\raggedright \footnotesize Frozen core (left) and all-electron (right) LiH STO-3G energies, computed with pUCCMCSD (orange circles - projected energy, green squares - variational energy) and tpUCCMCSD (blue circles - projected energy, red crosses - variational energy). In the left panel, the black line corresponds to UCCSD VQE. In the right figure, the black dashed line corresponds to the deterministic UCCSD benchmark of Cooper and Knowles\cite{Cooper2010} and the black solid line line to UCCSD VQE. In both cases, the stochastic energies agree with the VQE results to within $50\mu E\sr{h}$.}
    \label{fig:LiH-energy}
\end{figure}

In this section, we assess the viability of the tpUCCSD method as a screening technique for UCCSD-based VQE. Given one method is based on a projective approach, while the other is variational, we first ascertain whether the wavefunctions obtained by the two approaches are sufficiently similar for tpUCCSD amplitudes to be a good predictor of VQE UCCSD amplitudes. To assess this, we turn our attention to the LiH molecule. In the STO\nobreakdash-3G basis, this system consists of 4 electrons in 12 spin-orbitals, which can easily be treated by both VQE and tpUCCMCSD. The system can be further simplified by freezing the core Li $1s$ electrons. As seen in \cref{fig:LiH-energy}, in both cases the energies obtained by the two methods agree within $50\mu E\sr{h}$ for both the expectation value and the projected energy for tpUCCMCSD.

However, for tpUCCMCSD predictions to be useful as a starting guess or screening for VQE UCCSD amplitudes, the methods must agree not only in energy, but also in individual cluster amplitudes. In order for this agreement to be achieved, care must be taken that the same ordering is used in both ans\"atze. While for such small systems any arrangement of the excitors in the trotterized UCC wavefunction will be able to describe the ground state wavefunction, the individual amplitudes may vary extensively, particularly in the more highly correlated regime.\cite{supp}  For frozen core LiH, the tpUCCMCSD amplitudes agree with their deterministic counterparts within the $1\sigma$-error bars.

Having convinced ourselves that the VQE UCCSD and tpUCCMCSD results are compatible, we investigate the latter's potential as a screening technique for the former. \Cref{tab:lih-amp} gives the number of amplitudes in the LiH wavefunction of different orders of magnitude in a stochastic snapshot of the tpUCCMCSD expansion. We find that the lowest threshold ($t > 0.001$), recovers more than 99.9\%  of the correlation energy at all geometries, while decreasing the size of the considered Hilbert space by 14~-~43\%.
\onecolumngrid

\begin{table}[h]
    \centering
    \footnotesize
    \begin{tabular}{c|c|c|c|c|c|c|c}
         Bond length (\AA) & $\Sigma^+$ UCCSD Hilbert Space & $t > 0.1$ & \% $E_\mathrm{corr}$ & $ t > 0.01$ & \% $E_\mathrm{corr}$& $t > 0.001$& \% $E_\mathrm{corr}$ \\
         \hline
         0.995& 35 & 0 & 0 & 8 & 97.43& 29 & 99.93\\
         1.395& 35 & 1 & 71.20& 8 &98.75& 19& 99.94\\
         1.795& 35 & 1 & 70.54& 8 & 98.94 & 22& 99.95\\
         2.195& 35 & 3 & 88.90& 8 & 99.13& 25& 99.99\\
         \hline
    \end{tabular}
    \caption{\raggedright \footnotesize Size of the totally symmetric $\Sigma^+$ UCCSD Hilbert space and number of amplitudes above different thresholds for LiH in the STO-3G basis at a range of bond lengths, together with the percentage of the correlation energy recovered using each of the thresholds as a cutoff for amplitudes included in the VQE UCCSD ansatz, relative to the completer VQE UCCSD calculation.}
    \label{tab:lih-amp}
\end{table}

\twocolumngrid

\noindent  Even $t > 0.01$ generally recovers more than 97\% of the correlation energy in all cases, despite only using 8 parameters.

Following on from these results, we investigate the applicability of tpUCCMC screened VQE to a series of small molecules in different scenarios: \ce{CH2}, which has a triplet ground state, \ce{N2} which becomes increasingly strongly correlated as the bond between the nitrogen atoms is broken and \ce{N2H2}, where there is a crossing of diabatic states as it rotates from a \textit{trans} to a \textit{cis} geometry. In all cases, we use a snapshot of the amplitudes at the end of a short tpUCCMCSD run for screening.

\begin{figure}[h]
\centering
    \includegraphics[width=\textwidth, trim = 3cm 0 3cm 0, clip]{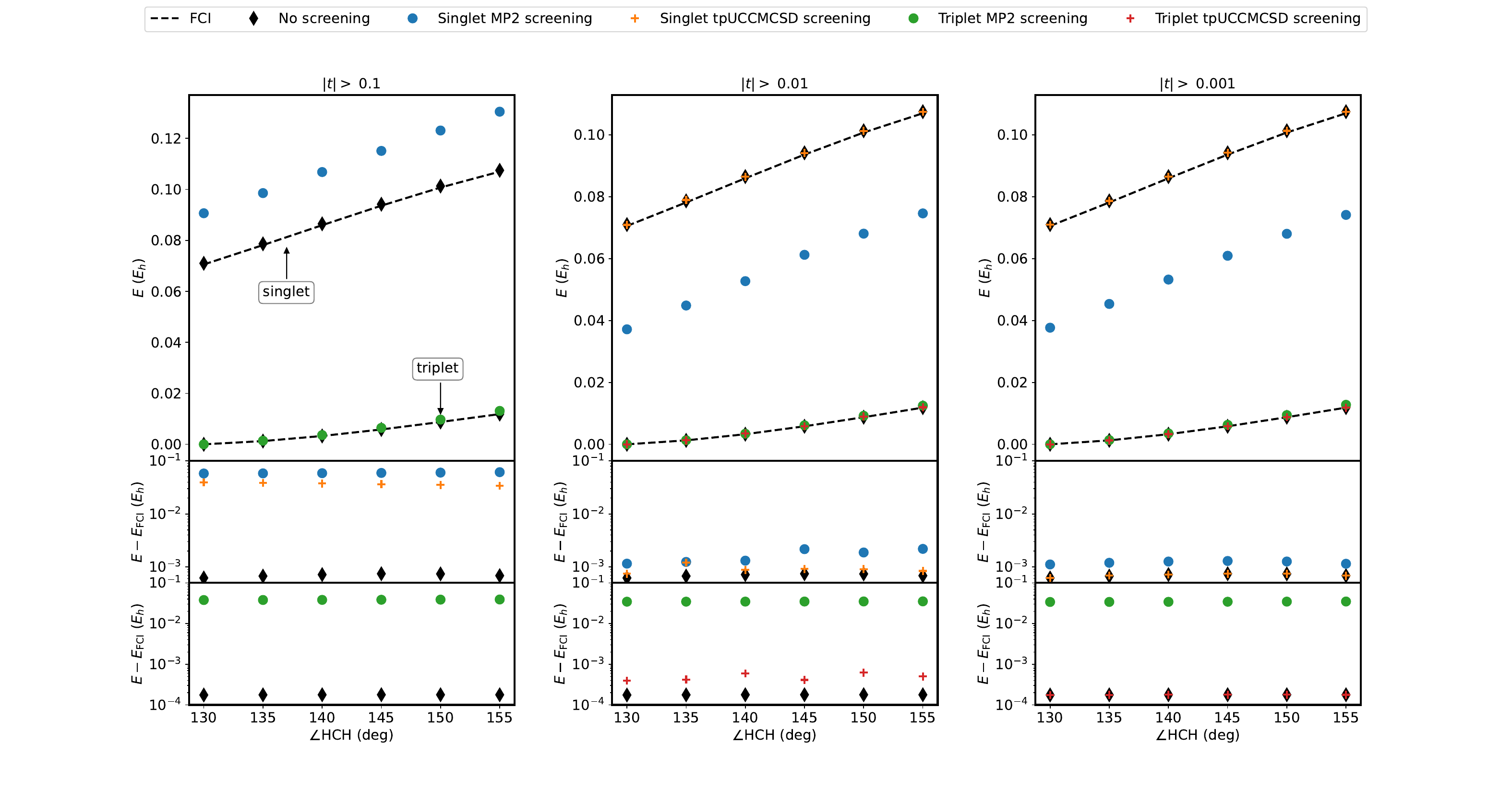} 
    \caption{\raggedright \footnotesize {\ce{CH2} singlet and triplet energy computed with UCCSD VQE across the bending mode at $r_\mathrm{CH} = 1.08$\AA with different coefficient thresholds, using MP2 (blue and green circles) and tpUCCMCSD (orange and red pluses) as screening methods. From left to right, we consider coefficients greater than 0.1, 0.01 and 0.001 respectively. All energies in the top panel are relative to the lowest computed energy along the binding curve obtained with the same method - in this case the triplet energy at $\angle HCH = 130^\circ$. For all methods, the error relative to FCI is given in the middle (for the singlet) and bottom (for the triplet) panels. In tpUCCMCSD screening, there are no amplitudes $|t| > 0.1$ for the triplet state.}}
    \label{fig:CH2-singlet-bend}
\end{figure}

\begin{figure}[h]
\centering
    \includegraphics[width=\textwidth, trim = 3cm 0 3cm 0, clip]{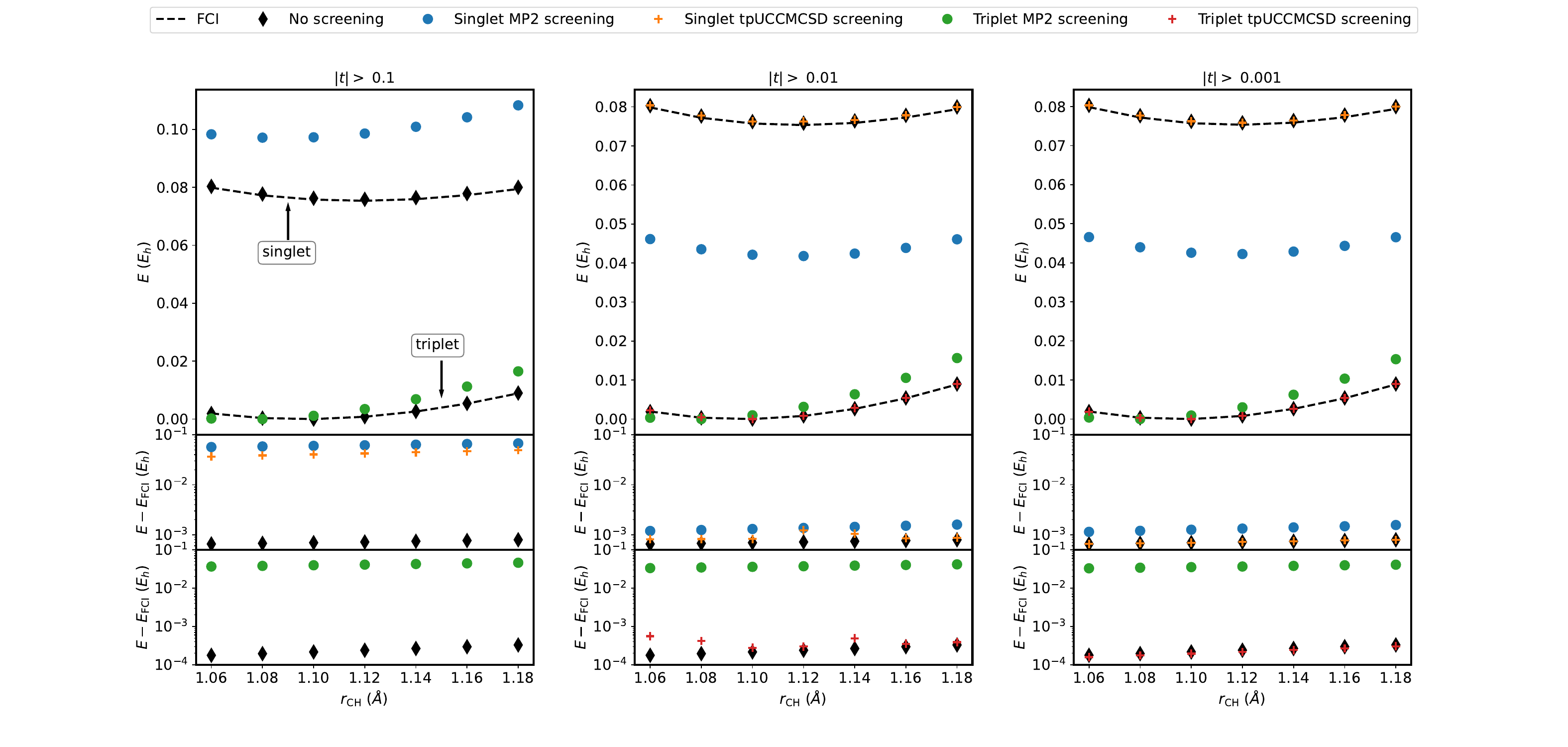} 
    \caption{\raggedright \footnotesize {\ce{CH2} singlet and triplet energy computed with UCCSD VQE across the stretching mode at $\angle HCH = 135^\circ$ with different coefficient thresholds, using MP2 (blue and green circles) and tpUCCMCSD (orange and red pluses) as screening methods. From left to right, we consider coefficients greater than 0.1, 0.01 and 0.001 respectively. All energies in the top panel are relative to the minimum energy along the binding curve obtained with the same method - in this case the triplet energy at $r_\mathrm{CH} = 1.08$\AA. For all methods, the error relative to FCI is given in the middle (for the singlet) and bottom (for the triplet) panels. In tpUCCMCSD screening, there are no amplitudes $|t| > 0.1$ for the triplet state.}}
    \label{fig:CH2-singlet-stretch}
\end{figure}

\begin{figure}[h!]
\centering
    \includegraphics[width=\textwidth, trim = 3cm 0 3cm 0, clip]{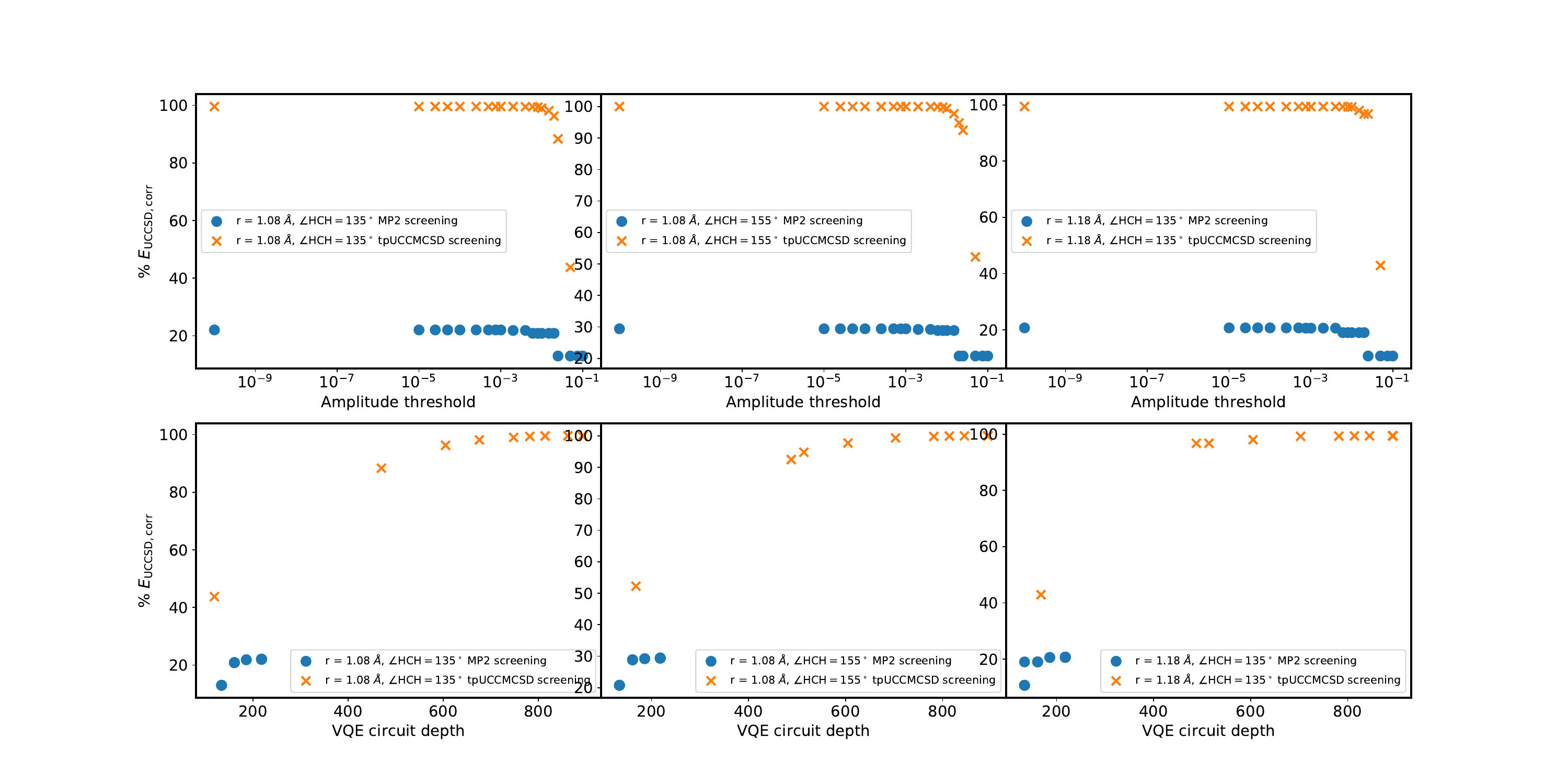} 
    \caption{\raggedright \footnotesize Percentage of UCCSD correlation energy recovered for triplet \ce{CH2} with MP2 (blue circles) and tpUCCMCSD (orange crosses) screening at a series of geometries. MP2 fails to converge for the triplet state, while tpUCCMCSD allows recovery of significant fractions of the correlation energy while reducing the circuit depth by up to half for all geometries.}
    \label{fig:CH2_depth}
\end{figure}

For \ce{CH2}, considered with frozen 1s electrons on the C atom, we once again observe good agreement between tpUCCMCSD and VQE energies and wavefunctions, provided the same excitor ordering is used throughout.\cite{supp} {In this study, we consider both the symmetric stretch at an angle $\angle HCH = 135^\circ$ and the bend with $r_\mathrm{CH} = 1.08$\AA, close to the experimental equilibrium geometry ($\angle \mathrm{HCH} = 133.9^\circ$, $r_\mathrm{CH} = 1.075$\AA)\cite{Jensen1982}} In this case, the tpUCCMCSD amplitudes are less spread out in magnitude and therefore the screening is less efficient at decreasing the number of VQE parameters to be considered (see \cref{tab:ch2-amp} for an example). As expected from this distribution of amplitudes, screening at either $t > 0.01$ or $t > 0.001$ gives errors of less than 1 milliHartree relative to the full UCCSD VQE calculation, for both singlet and triplet \ce{CH2}. Using MP2 as a screening method is competitive for the singlet state (see \cref{fig:CH2-singlet-bend,fig:CH2-singlet-stretch}), while for the triplet, the MP2 screened results exhibit a systematic error of more than 10 milliHartree which persists across all screening thresholds. {Even when considering relative energies, MP2 screened calculations overestimate the energy of the triplet as the C-H bonds are stretched, as seen in \cref{fig:CH2-singlet-stretch} and generally underestimate the singlet-triplet gap. Considering quantum resources, we note that the tpUCCMC screening described above leads to significant depth reduction with relatively little correlation loss, as can be seen in \cref{fig:CH2_depth}, which also highlights the unsuitable quality of MP2-screened results independently of circuit depth.}

The N$_2$ molecule as it approaches dissociation is one of the archetypal examples of static correlation and poses significant challenges to methods such as coupled cluster, which are based around the assumption that a single determinant dominates the wavefunction. For example, for $r_\mathrm{NN} > 3.6 a_0$ CCSD(T)\cite{Raghavachari1989} overestimates the  correlation 
\onecolumngrid


\twocolumngrid

\begin{figure*}
    \centering
    \includegraphics[width=\textwidth, trim = 3cm 0 3cm 0, clip]{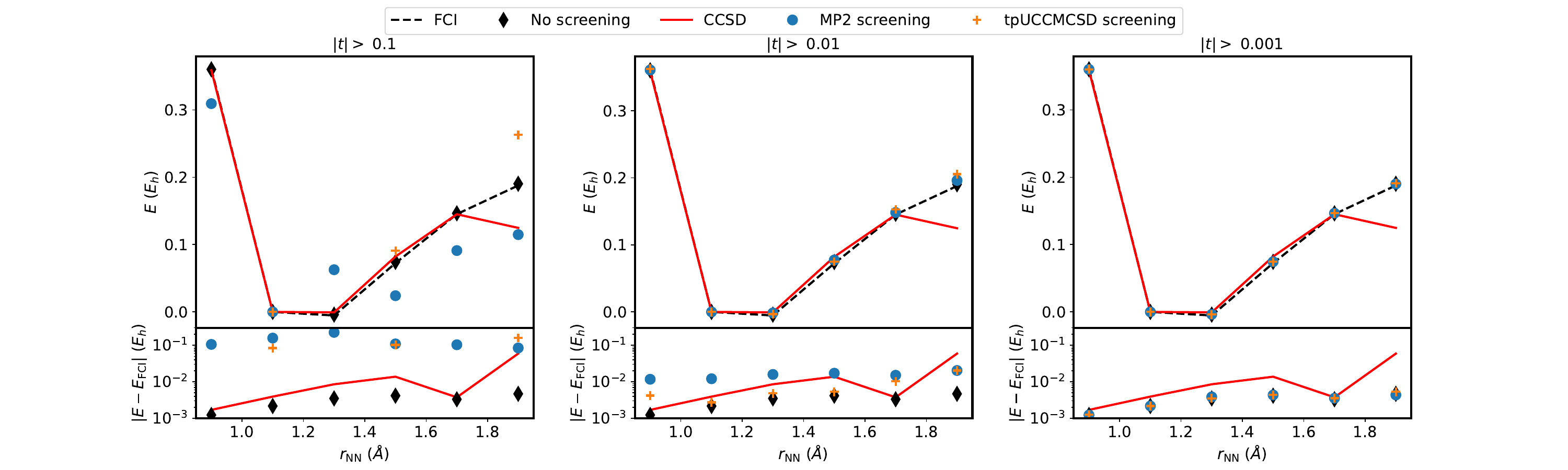}
    \caption{\raggedright \footnotesize \ce{N2} frozen-core binding curve computed with UCCSD VQE with different coefficient thresholds, using MP2 (blue cicles) and tpUCCMCSD (orange crosses) screening. From left to right, we consider coefficients greater than 0.1, 0.01 and 0.001 respectively. {All energies in the top panel are relative to the energy at $r_\mathrm{NN} = 1.1$\AA\ obtained with the same method.} Energies converge to the unscreened UCCSD VQE value (solid black line) with number of parameters for both methods. At intermediate screening, tpUCCMCSD performs better than MP2 at compressed bond lengths. CCSD energies (red line) are show for comparison.}
    \label{fig:N2-energy}
\end{figure*}

\begin{figure}[h!]
\centering
    \includegraphics[width=0.5\textwidth, trim = 1cm 1cm 2cm 1.5cm, clip]{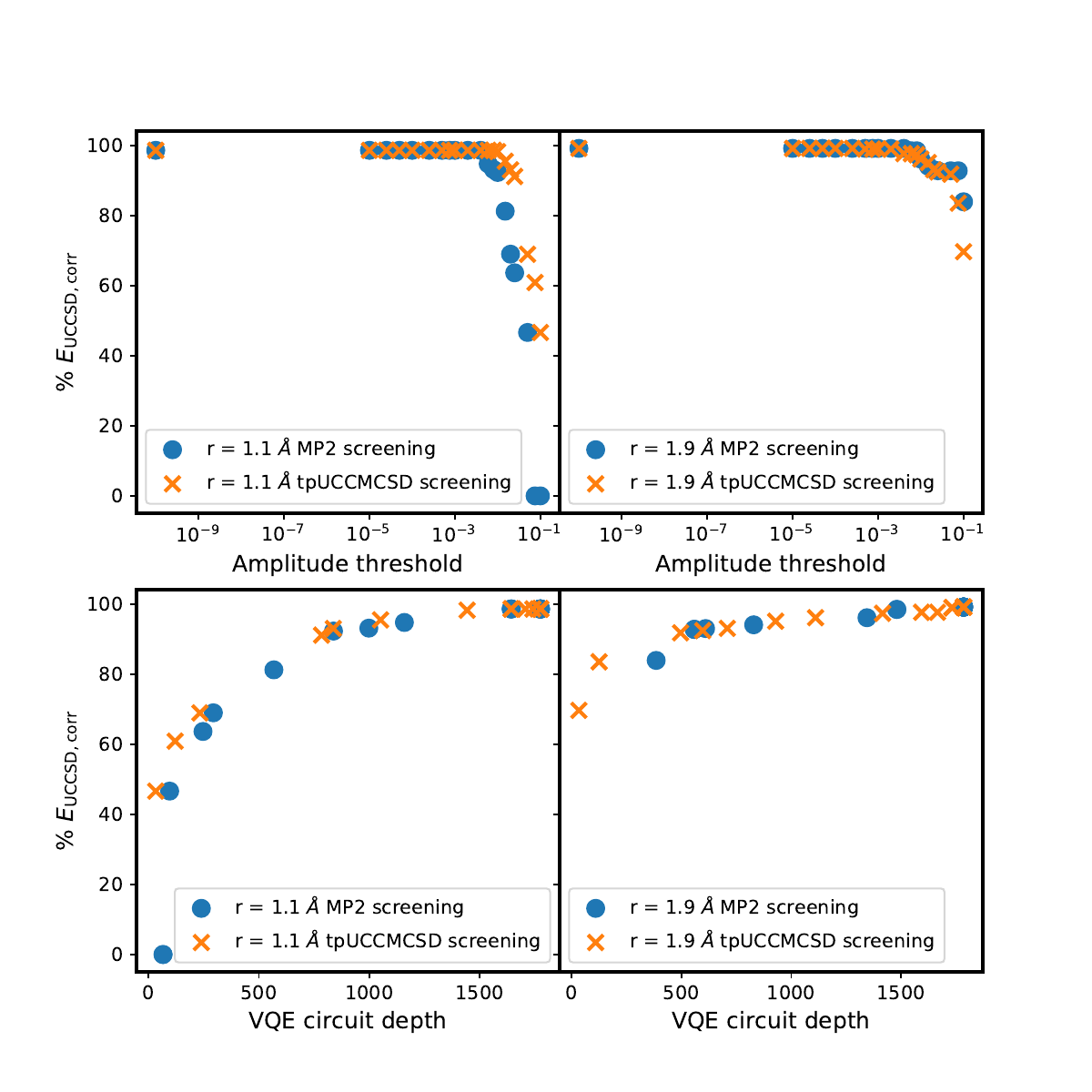} 
    \caption{\raggedright \footnotesize Percentage of UCCSD correlation energy recovered for \ce{N2} with MP2 (blue circles) and tpUCCMCSD (orange crosses) screening near equilibrium (left) and for a stretched bond (right). Performance is comparable between the two methods, with tpUCCMC exhibiting slightly faster convergence in terms of both threshold and depth.}
    \label{fig:N2_depth}
\end{figure}

\begin{table}[h]
    \centering
    \footnotesize
    \begin{tabular}{c|c|c|c|c}
         Bond length (\AA) & UCCSD Hilbert Space & $t > 0.1$ & $t > 0.01$ & $t > 0.001$ \\
         \hline
         1.06& 31 & 0 & 24 & 30 \\
         1.08& 31 & 0 & 25 & 30 \\
         1.10& 31 & 0 & 26 & 29 \\
         1.12& 31 & 0 & 26 & 29 \\
         1.14& 31 & 0 & 25 & 29 \\
         1.16& 31 & 0 & 26 & 30 \\
         1.18& 31 & 0 & 26 & 29 \\
         \hline
    \end{tabular}
    \caption{\raggedright \footnotesize Size of the totally symmetric UCCSD Hilbert space and number of amplitudes above different thresholds for triplet \ce{CH2} in STO-3G at a range of CH bond lengths, with $\angle\mathrm{HCH} = 135 ^\circ$.}
    \label{tab:ch2-amp}
    \end{table}
\noindent energy and diverges as the bond length is increased.\cite{Chan2004} As can be seen in \cref{fig:N2-energy}, the unitary, variationally optimised ansatz avoids the failure of the traditional CCSD ansatz, with the energy consistently above the FCI value and generally lower errors. In this case, both MP2 and tpUCCMCSD screening provide similar quality results, but the system highlights a few interesting differences between the two methods. First, in the left panel of \cref{fig:N2-energy} we observe that, while in MP2 there are consistently some amplitudes allowed by the screening, tpUCCMCSD does not have any amplitudes above 0.1 for half of the geometries considered. As such, the screened method cannot provide any improvement over Hartree--Fock. Obviously, this threshold is too high and the results obtained with either MP2 or tpUCCMCSD screening all have significant {systematic and non-parallelity} errors (10 - 500 milliHartree), making it inappropriate for the treatment of this system. Secondly, in the previous examples we have generally found that tpUCCMCSD screening converges slightly faster with screening threshold than MP2, or indeed MP2 fails to converge at all. In this case, we find that tpUCCMCSD converges faster than MP2 for compressed bonds, but they are comparable in the stretched regime, {as is further highlighted in \cref{fig:N2_depth}, which shows that once again screening can be used to meaningfully reduce circuit depth without significant accuracy loss.} The behaviour can be correlated to the number of amplitudes in the ansatz at each geometry, with tpUCCMCSD including more parameters at shorter bond lengths than MP2 (see \cref{tab:n2-amp}). Both of these features emphasize the fact that a constant amplitude threshold does not necessarily generate a consistent number of parameters across all geometries of a given system. Where such consistency is desirable, which will often be the case when constructing a description of a physical system, a threshold based directly on the number of parameters included may be preferable{,although it is not without its own challenges, as we discuss below.}

\begin{figure*}
    \centering
    \includegraphics[width = \textwidth, trim = 3cm 1cm 3cm 0, clip]{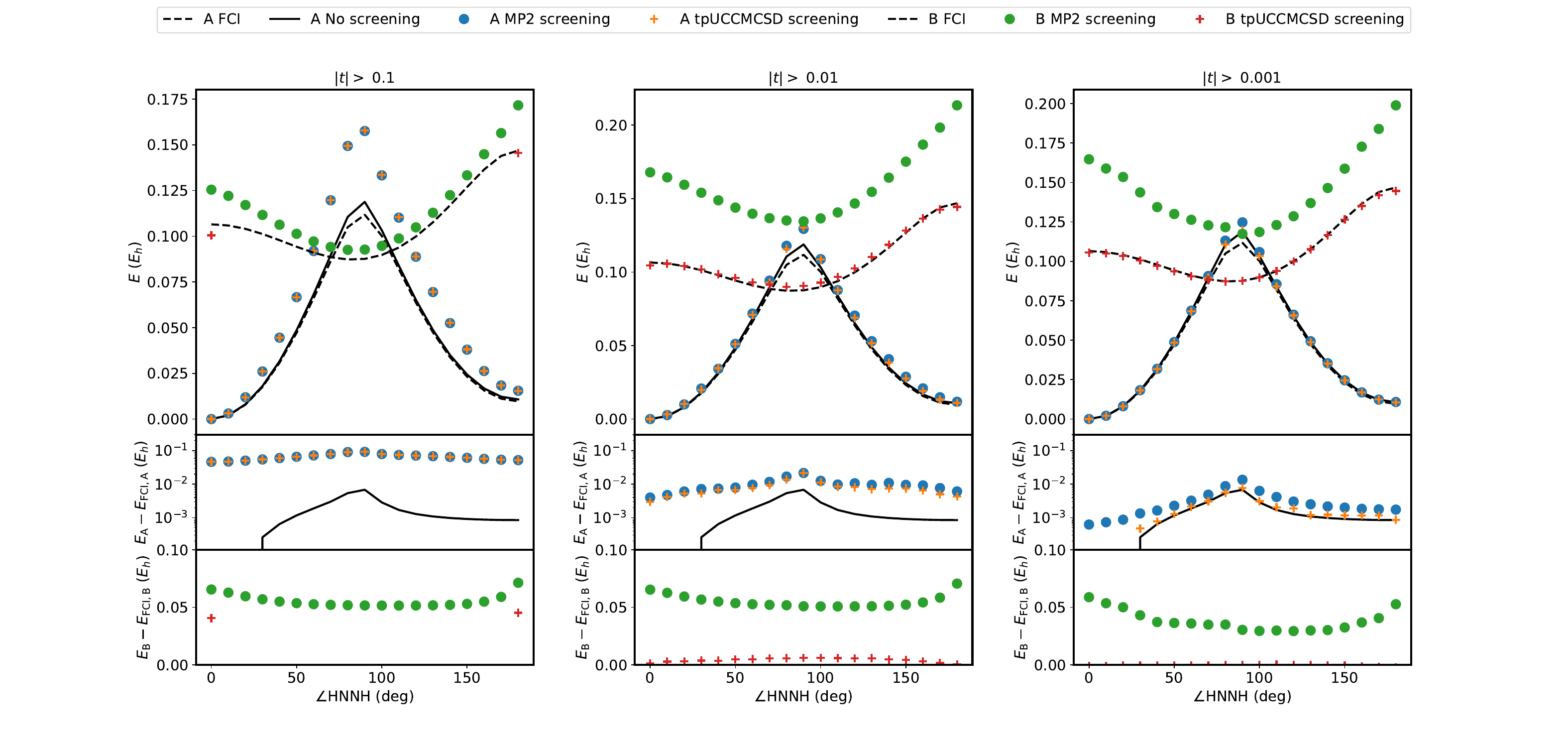}
    \caption{\raggedright \footnotesize \ce{N2H2} energy computed with UCCSD VQE with different coefficient thresholds, using MP2 (blue and green circles) and tpUCCMCSD (orange and red crosses) for screening. We consider two states, a closed-shell singlet state of A symmetry in the C2 point group and a triplet state of B symmetry. The A state is the ground state around the cis and trans geometries of \ce{N2H2}, but the triplet becomes favoured around a rotation angle of 90$^\circ$. From left to right, we consider coefficients greater than 0.1, 0.01 and 0.001 respectively. {In all cases, the energies in the top panel are quoted relative to the energy of the singlet state in the \textit{trans} geometry, computed with the same method.} For the A state, as with all singlet molecules so far, both MP2 and tpUCCMCSD converge towards the true UCCSD energy, although in this case at 0.001 MP2 is noticeably worse than tpUCCMCSD. For the B state however, MP2 is unable to capture the correlation, and does not improve with added amplitudes. {This gives unphysical results, with the B state higher in energy than the A state for all geometries except $\angle\mathrm{HNNH} = 90^\circ$ even at the lowest threshold considered.}}
    \label{fig:N2H2-energy}
\end{figure*}

Finally, the energy landscape of the \textit{cis}--\textit{trans} interconversion of N$_2$H$_2$ allows us to investigate another highly-correlated regime around the transition state of this transformation. For simplicity, we investigate the rotation about the {nitrogen--nitrogen} bond starting from the equilibrium geometry of \textit{trans} \ce{N2H2}\cite{Demaison1997}, without allowing any relaxation of other geometrical parameters. At the Hartree--Fock level, the singlet state of this system would be characterised by two diabatic states which cross at $86.8(1)^\circ$\cite{supp}. This introduces an unphysical discontinuity in the energy surface, which often remains even after the application of further correlated methods.\cite{Mach1998} In order to be able to treat this system using VQE, we freeze the 8 lowest-energy electrons. The resulting system exhibits an additional interesting feature as two FCI states of different spatial symmetry cross, as can be seen in \Cref{fig:N2H2-energy}. The conservation of various symmetries --- spin, particle number, point group --- in quantum circuits is an interesting problem and in this system it turns out to be crucial for the correct description of the energy surface. Not enforcing point group symmetry leads to a very compelling yet unphysical state which transitions smoothly from the A to the B state as the system approaches a $90^\circ$ rotation. 

Once point group symmetry constraints are considered, VQE based on the UCCSD ansatz with the excitations symmetry filtered correctly reproduces the behaviour observed in the exact surface, however starting from an $m_s = 0$ open-shell Hartree--Fock reference or an $m_s = 1$ reference leads to different energies for the B state. For the rest of our analysis we focus on the $m_s = 0$ A state and the $m_s = 1$ B state, which correspond to the two lowest-energy states in the system. The order of magnitude of amplitudes in a short tpUCCMCSD run for this system at various rotation angles are given in \cref{tab:n2h2-amp}. This system exhibits many of the behaviours we noted in smaller examples. For the singlet state, both MP2 and tpUCCMCSD screened UCCSD VQE converge to the correct UCCSD energy with increasing number of amplitudes, but for the triplet state MP2 fails to converge, retaining an error of at least 30 milliHartree even for the lowest threshold - see \cref{fig:N2H2-energy}. {As there is no systematic error in the singlet state, this causes MP2 screening to predict the wrong ordering of the singlet and triplet states for rotation angles around $90^\circ$}.  

\begin{figure}
    \centering
    \includegraphics[width=0.5\textwidth]{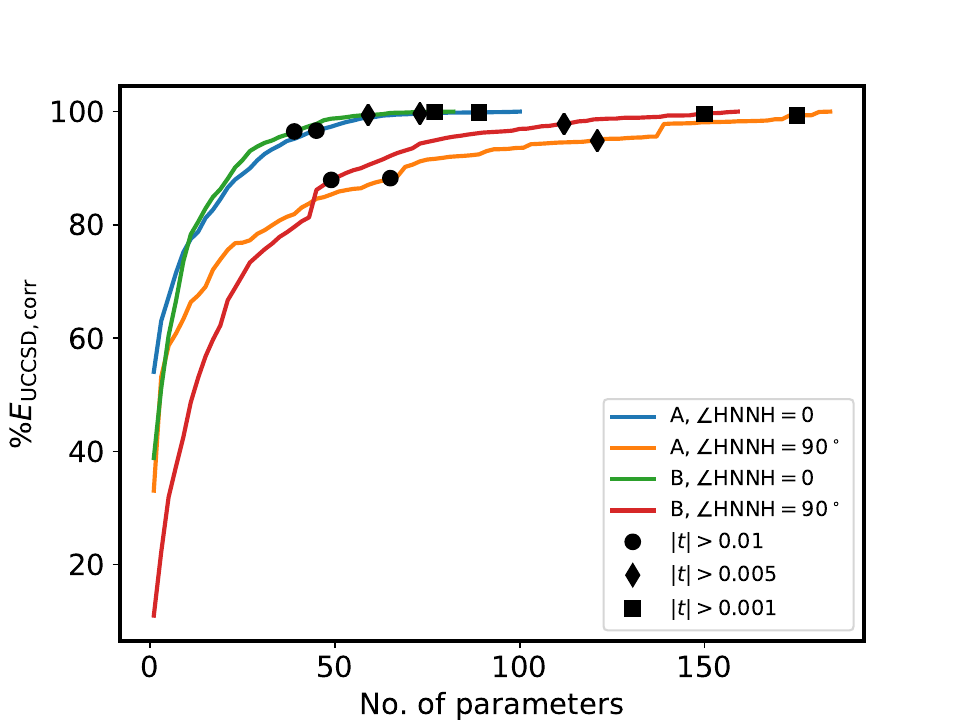}
    \caption{\raggedright {\footnotesize Convergence of the correlation energy with number of UCCSD amplitudes in the lowest-lying electronic states of \ce{N2H2} at $\angle\mathrm{HNNH} = 0$ and $\angle\mathrm{HNNH} = 90^\circ$. The percentage of the energy recovered monotonically increases with parameter number, but a large fraction is encoded within the first few parameters to be included. Behaviour is similar between the two states at matching geometries, dominated by the difference in the maximum number of parameters rather than the different degrees of correlation character in the two states. Some corresponding amplitude thresholds are marked for comparison.}}
    \label{fig:constant_cutoff}
\end{figure}

\begin{figure} [h]
    \centering
    \includegraphics[width = 0.5\textwidth, trim = 1cm 1cm 2cm 1.5cm, clip]{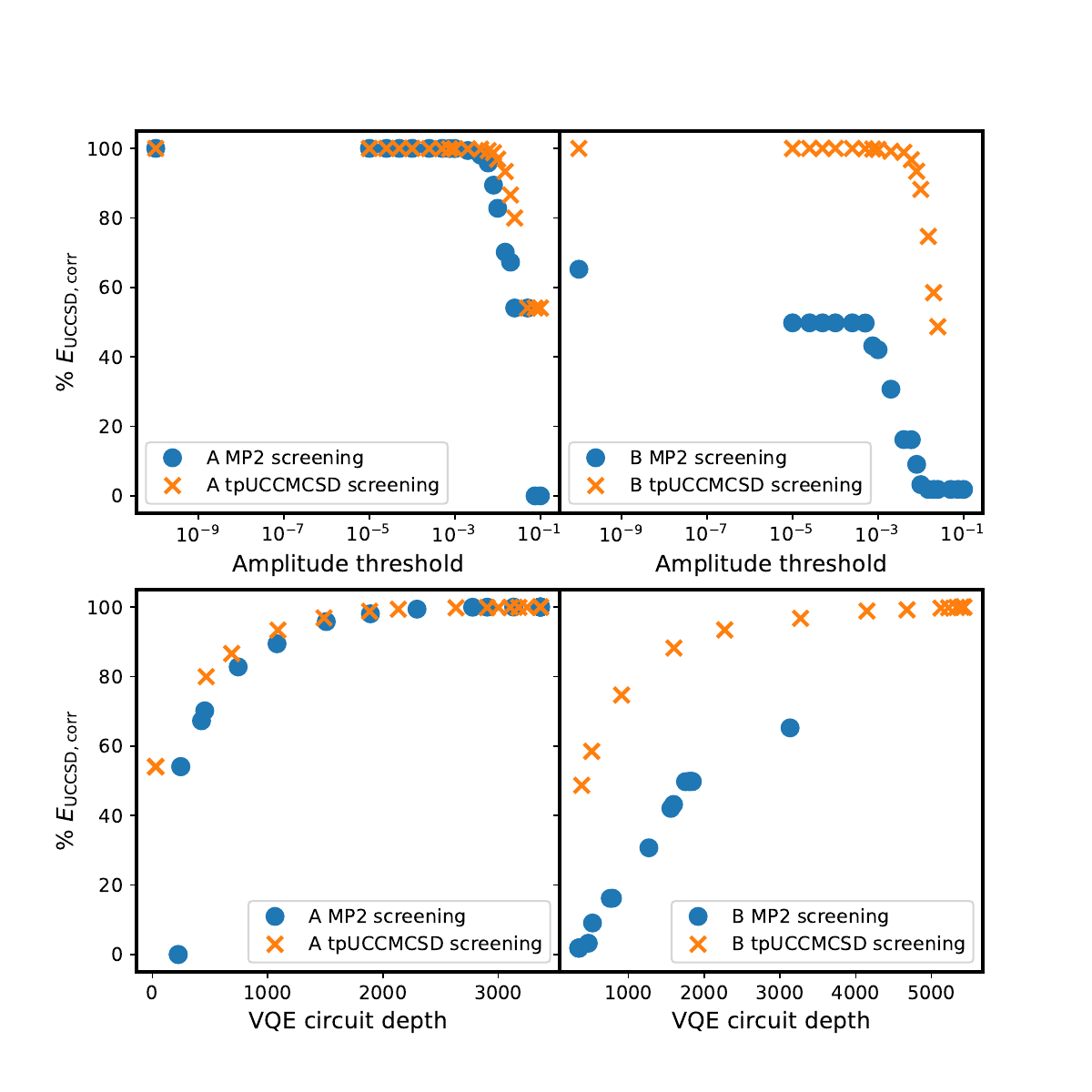}
    \caption{\raggedright \footnotesize \ce{N2H2} percentage of UCCSD correlation energy recovered with MP2 (blue circles) and tpUCCMCSD (orange crosses) screening for the A state at the \textit{trans geometry} (left) and the B state at $\angle\mathrm{HCH}=90^\circ$ (right) as a function of screening threshold (top) and circuit depth (bottom). tpUCCMCSD outperforms MP2 over the entire range of screening thresholds and for the triplet B state any MP2 screened state fails to converge to the correct energy.}
    \label{fig:N2H2-depth}
\end{figure}

\cref{fig:N2H2-depth} shows the convergence of VQE UCCSD energy as a function of amplitude threshold and circuit depth for the A state at $0^\circ$ and the B state at $90 ^\circ$. This highlights the same trend as before, with tpUCCMCSD screening only slightly more efficient than MP2 as a function of screening threshold for the singlet state, but clearly superior for the triplet. In both cases, using tpUCCMCSD it is possible to recover more than 90\% of the correlation energy with a circuit that is only half as deep as the full VQE UCCSD implementation. {In \cref{fig:constant_cutoff} we consider the correlation energy for the \textit{trans} geometry of \ce{N2H2} and the transition state at $90^\circ$ directly as a function of the number of parameters in the UCCSD expansion, using the truncation order from tpUCCMCSD. Due to the change in point-group symmetry from C$_2h$(C$_2v$) at the trans(cis)-geometry to C$_2$ along the rest of the binding curve, the number of possible parameters is halved at the ends. Therefore, where the symmetry of the system changes, computations based on a constant number of amplitudes are non-trivial to implement.}
\onecolumngrid

\begin{table}[h]
    \centering
    \footnotesize
    \begin{tabular}{c|c|c|c|c|c|c|c}
         Bond length (\AA) & UCCSD &\multicolumn{3}{c}{tpUCCMCSD} \vline &\multicolumn{3}{c}{MP2} \\
         &                 Hilbert space & $t >0.1$ & $t >0.01$ & $t >0.001$ & $t >0.1$ & $t >0.01$ & $t >0.001$ \\
         \hline
         0.9& 54 & 0 & 36 & 52 & 2& 24 & 53\\
         1.1& 54 & 2 & 44& 52 & 2 & 26& 49\\
         1.3& 54 & 0 & 42& 53 & 2 & 30& 49\\
         1.5& 54 & 3 & 47& 53 & 4 & 35& 53\\
         1.7& 54 & 0 & 42& 53 & 8& 39& 53\\
         1.9& 54 & 2 & 34& 52 & 12& 41& 53\\
         \hline
    \end{tabular}
    \caption{\raggedright \footnotesize Size of the totally symmetric UCCSD Hilbert space and number of amplitudes above different thresholds for frozen core \ce{N2} in STO-3G at a range of bond lengths, screened with both MP2 and tpUCCMCSD.}
    \label{tab:n2-amp}
\end{table}

\twocolumngrid

{As a final point, we consider the scaling of such screening methods with basis set size. While the VQE method, even simulated on a classical machine, is currently limited to a minimal basis for the systems considered here, the sparse tpUCCMC approach can tackle more basis functions. Given results from this method have been shown to agree well with VQE, we use them to predict its expected behaviour. We consider the \ce{N2} molecule in the cc-pVDZ, cc-pVTZ and cc-pVQZ basis sets. We start by running a tpUCCMCSD calculation for these systems to convergence. We then truncate the resulting parameter sets and rerun the calculation in this limited Hilbert space. Results are given in \cref{tab:N2_bases}. In all cases, the correlation energy increases with the number of amplitudes included, but we find that more than 85\% of it is accounted for at a threshold of $|t| > 0.001$ in all cases. While at this value the number of amplitudes required increases with basis set size, the proportion of the Hilbert space decreases, suggesting larger relative resource reductions are possible for larger basis sets.}

\section{Conclusion}
We have shown that UCCMC can be used successfully as a screening method for UCCSD-based VQE, with short, potentially unconverged stochastic runs providing a set of amplitudes that may be truncated at a given coefficient threshold before being further optimised by VQE. We expect the method would also be compatible with the newly developed Projected Quantum Eigensolver.\cite{Stair2021} Importantly, the classical screening approach is entirely a pre-processing step, so requires no additional quantum resources. 

Standard noise models such as amplitude and phase damping\cite{Nielsen2011} predict exponential decay of the quantum signal with time, controlled by the relaxation time $T_1$ and dephasing time $T_2$\cite{Ghosh2012}. Significantly reducing the depth, and therefore the computation time, of the VQE circuit, as we have seen is possible for \ce{N2H2}, where more than 90\% of the correlation energy could be recovered with half the circuit depth, would therefore lead to a major reduction in the error rates observed on real hardware. {Furthermore, two-qubit gates such as CNOTs are responsible for most gate noise on current devices. The number of such gates is reduced proportionally to the number of parameters removed in screened calculations. Errors in the single qubit gates should also be reduced, as larger values for the angles in the $R_z$ gates involved in the UCC \textit{ansatz} are easier to implement with high fidelity than lower values.}

\begin{table}[h]
    \centering
    \footnotesize
    \begin{tabular}{c|c|c|c|c|c}
         State & \shortstack{Dihedral\\ angle (deg)} & \shortstack{UCCSD \\ Hilbert Space} & $t > 0.1$ & $t > 0.01$ & $t > 0.001$ \\
         \hline
         \multirow{7}{*}{A} & 0  & 100 & 1 & 46 & 90 \\
         &30 & 185 & 1 & 60 & 170 \\
         &60 & 185 & 1 & 71 & 166 \\
         &90 & 185 & 1 & 65 & 175 \\
         &120& 185 & 1 & 65 & 161 \\
         &150& 185 & 1 & 56 & 172 \\
         &180& 100 & 1 & 42 & 97 \\
         \hline
        \multirow{7}{*}{B} & 0  & 83 & 1 & 40 & 77 \\
        & 30 & 160 & 0 & 44 & 140 \\
        & 60 & 160 & 0 & 51 & 147 \\
        & 90 & 160 & 0 & 49 & 150 \\
        & 120& 160 & 0 & 51 & 149 \\
        & 150& 160 & 0 & 45 & 147 \\
        & 180& 83 & 1 & 36 & 76 \\
         \hline
    \end{tabular}
    \caption{\raggedright \footnotesize Size of the UCCSD Hilbert space and number of tpUCCMCSD amplitudes above different thresholds for the lowest singlet and triplet \ce{N2H2} states in STO-3G at a range of dihedral $\angle \mathrm{HNNH}$.}
    \label{tab:n2h2-amp}
    \end{table}

{We have compared tpUCCMC screening with the pre-existent MP2-based screening approach. In the case of singlet states, the results using UCCMC for screening are seen to converge slightly faster with screening threshold than the corresponding MP2-screened values. For triplets, we have found that the MP2-screened calculations fail to converge to the true ground state at all. This highlights the fact that tpUCCMC screening can be easily applied to any molecular system, while more care is needed when considering whether MP2 is suitable for a particular application. Furthermore, tUCC amplitudes are often highly dependent on operator ordering in the cluster expansion. While tpUCCMC can be adapted to match any operator ordering, MP2 can only generate one set of amplitudes, which may be better or worse predictors for the corresponding tUCC values depending on the chosen order.} For all molecules we have considered, the accuracy of the results increases monotonically with the number of amplitudes included in the ansatz, allowing one to balance result quality with resource limitations as necessary.

{tpUCCMCSD results for N$_2$ in increasingly large basis sets confirm our expectation that the benefits of such screening methods get more significant as one considers larger systems.} While the Hilbert spaces required for UCC calculations will scale with high-order polynomials of the system size, {a smaller fraction of} the amplitudes in this space will have large coefficients and therefore contribute significantly to the energy, at least in the largely single-reference, dynamically correlated areas of the landscape where coupled cluster methods perform well. Furthermore, it has been shown that linear scaling coupled cluster can be achieved by careful screening of amplitudes by distance\cite{Scuseria1999,Schutz2001,Flocke2004,Subotnik2006,Riplinger2013}.  {We expect that the trotterized UCC approach and its stochastic counterpart will, in principle, be able to take advantage of this, especially on systems of multiple molecules and propagate these gains forward into screened VQE calculations.}

\begin{table*}[]
    \centering
    {\footnotesize\begin{tabular}{c|c|c|c|c|c|c|c|c|c|c|c|c|c|c|c|c}
    \multirow{2}{*}{Basis set}  &  \multicolumn{2}{c}{$t > 0.1$} \vline &\multicolumn{2}{c}{$t > 0.05$}  \vline & \multicolumn{2}{c}{$t > 0.01$}  \vline  \vline & \multicolumn{2}{c}{$t > 0.005$}  \vline & \multicolumn{2}{c}{$t > 0.001$}  \vline & \multicolumn{2}{c}{$t > 0.0005$}  \vline & \multicolumn{2}{c}{$t > 0.0001$}  \vline & \multicolumn{2}{c}{All $t$}  \\
    \cline{2-17}
    & $E_\mathrm{corr}$ & $N_\mathrm{p}$  & $E_\mathrm{corr}$ & $N_\mathrm{p}$ & $E_\mathrm{corr}$ & $N_\mathrm{p}$ & $E_\mathrm{corr}$ & $N_\mathrm{p}$ & $E_\mathrm{corr}$ & $N_\mathrm{p}$ & $E_\mathrm{corr}$ & $N_\mathrm{p}$ & $E_\mathrm{corr}$ & $N_\mathrm{p}$ & $E_\mathrm{corr}$ & $N_\mathrm{p}$\\
    \hline
    cc-pVDZ & -0.0799(6) & 3 & -0.0980(5) & 5 & -0.260(1) & 205 & -0.3272(4) & 607 & -0.3570(7) & 1547 & -0.3592(5)& 2047 & -0.359(1)& 3245 &-0.3608(3)& 4256\\
    cc-pVTZ & -0.073(3)& 3 & -0.089(2) & 5 & -0.2413(6) & 226 & -0.300(1)& 611 &-0.4127(5) & 4544 &-0.422(2) & 7489 & -0.4367(3) & 15701 & -0.437(1) &26420\\
    cc-pVQZ &-0.034(1) & 2 &-0.091(1) & 8 &-0.217(2) & 239 &-0.279(2) & 762 &-0.426(3) & 5897 & -0.461(4) & 12413 & - 0.483(2)& 43345 & -0.493(6) &97994\\
    \hline
    \end{tabular}}
    \caption{\raggedright\footnotesize{Screened \ce{N2} UCCMCSD correlation energy $E_\mathrm{corr}$ and number of parameters $N_\mathrm{p}$ as a function of basis set and screening threshold. In all cases, more than 85\% of the correlation energy can be recovered using a threshold of $|t| > 0.001$, which corresponds to a decreasing proportion of the Hilbert space as the size of the basis increases. In the cc-pVQZ basis set, 98\% of the correlation energy can be recovered with only 44\% of parameters.}}
    \label{tab:N2_bases}
\end{table*}

\begin{acknowledgements}
M-A.F. is grateful to the Cambridge Trust and Corpus
Christi College for a studentship and A.J.W.T. to the
Royal Society for a University Research Fellowship under
Grant No. UF160398. The VQE numerical simulations in this work  were  performed  on  Microsoft  Azure  Virtual  Machines provided by the program Microsoft for Startups.
\end{acknowledgements}

\bibliography{./biblio}

\end{document}


\title[]{Supplemental Material for ``Reducing Unitary Coupled Cluster Circuit Depth by Classical Stochastic Amplitude Pre-Screening"}
\author{Maria-Andreea Filip}
    \affiliation{Yusuf Hamied Department of Chemistry\\Lensfield Road, CB2 1EW Cambridge\\ United Kingdom}
\author{Nathan Fitzpatrick}
    \affiliation{Cambridge Quantum Computing Ltd.\\ 13-15 Hills Road, CB2 1NL Cambridge\\ United Kingdom}
\author{David Mu\~noz Ramo}
    \affiliation{Cambridge Quantum Computing Ltd.\\ 13-15 Hills Road, CB2 1NL Cambridge\\ United Kingdom}
\author{Alex J. W. Thom}
    \affiliation{Yusuf Hamied Department of Chemistry\\Lensfield Road, CB2 1EW Cambridge\\ United Kingdom}
\maketitle

\onecolumngrid

\section{LiH amplitudes}

\begin{figure}[h!]
    \includegraphics[width=\textwidth]{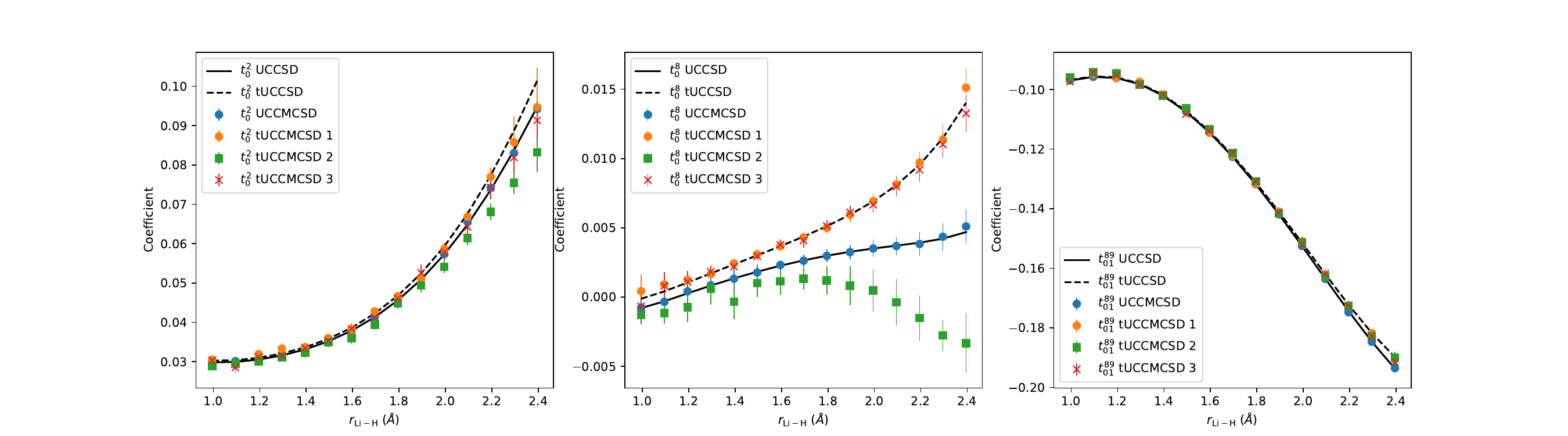}
    \caption{Frozen core LiH amplitudes for full pUCCMCSD and trotterized pUCCMCSD with 3 different orderings. The energies obtained agree in all cases, but the amplitudes themselves are significantly order-dependent. The Monte Carlo amplitudes are compared to deterministic pUCCSD and tpUCCSD (same as order 1) amplitudes.}
    \label{fig:LiH-coeffs}
\end{figure}
\twocolumngrid
\FloatBarrier

\onecolumngrid

\pagebreak
\section{\ce{CH2} amplitudes}
\begin{figure}[h!]
\centering
    \includegraphics[width=\textwidth]{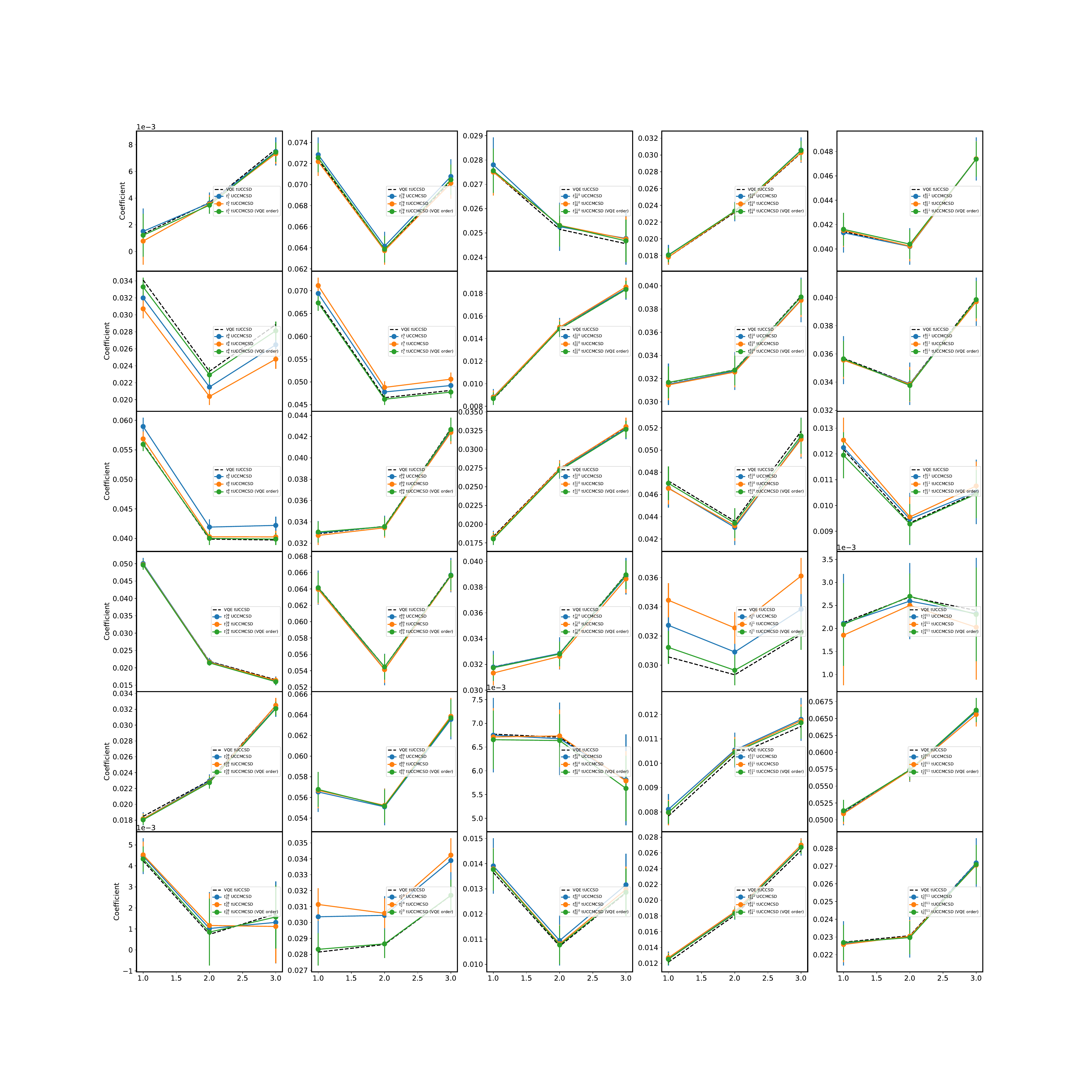} 
    \caption{Triplet \ce{CH2} amplitudes going from an enlarged HCH angle, to the equilibrium geometry to a symmetrically stretched geometry. The black line is VQE tUCCSD value and the green line corresponds to same ordering of excitors in tpUCCMCSD.  The blue and orange lines are pUCCMCSD and tpUCCMCSD using the default ordering\cite{Filip2020} respectively.}
    \label{fig:CH2-triplet}
\end{figure}
\twocolumngrid
\FloatBarrier

\onecolumngrid

\pagebreak
\section{\ce{N2H2} diabatic states}
\begin{figure}[h!]
\centering
    \includegraphics[width=\textwidth]{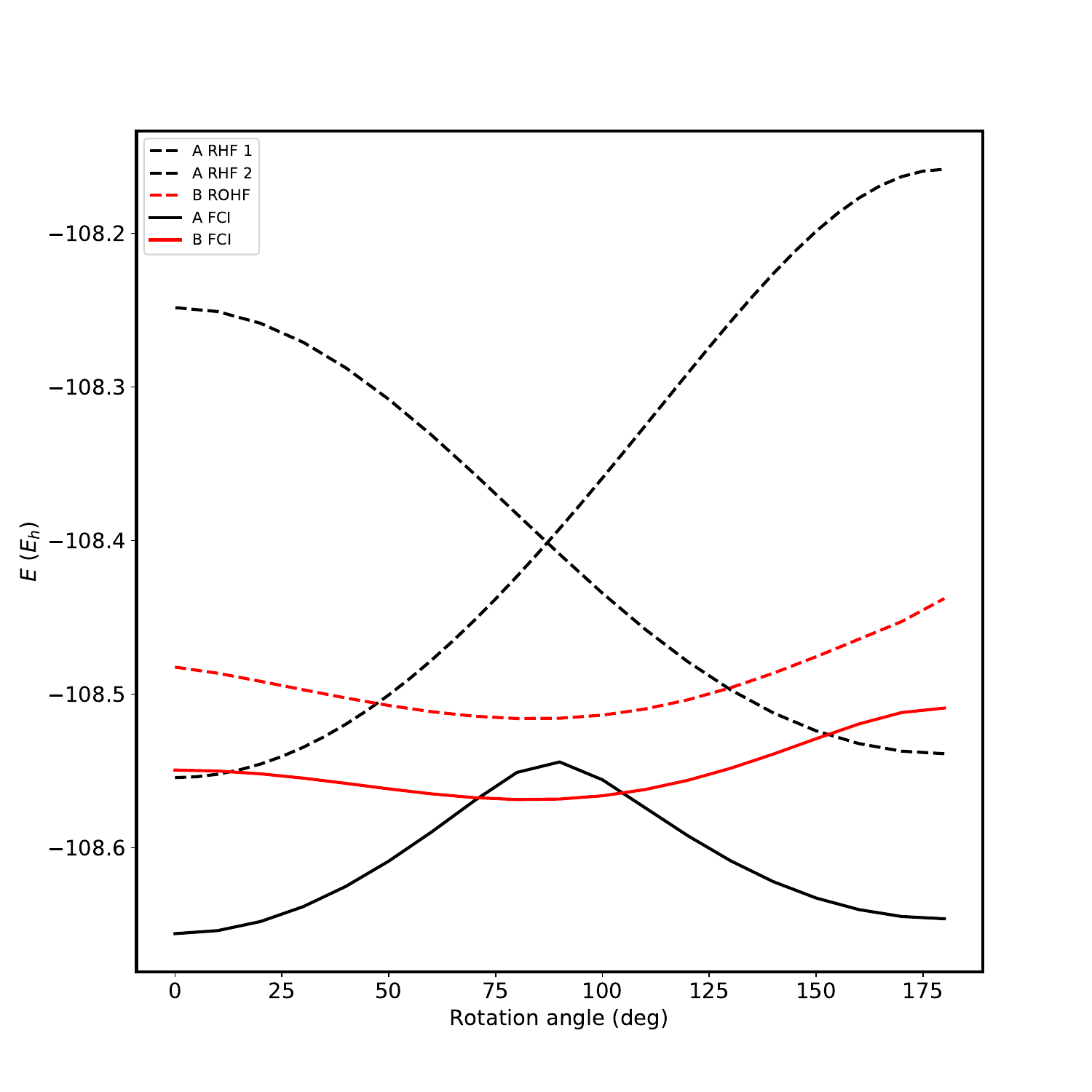} 
    \caption{Hartree-Fock and FCI energies for the A ($m_s = 0$) and B ($m_s = 1$) states of \ce{N2H2} in STO-3G as it rotates from a \textit{trans} to a \textit{cis} geometry.}
    \label{fig:N2H2-diabat}
\end{figure}

\FloatBarrier
\bibliography{./biblio}